\documentclass[runningheads]{llncs}

\usepackage[T1]{fontenc}

\usepackage{graphicx}
\usepackage[colorlinks=true, allcolors=blue]{hyperref}
\usepackage{color}

\usepackage{amsmath,amsfonts,amssymb}
\usepackage{graphicx}
\usepackage{subcaption}

\begin{document} 

\title{What Does Normal Even Mean? Evaluating Benign Traffic in Intrusion Detection Datasets}
\titlerunning{What Does Normal Even Mean}

\author{Meghan Wilkinson\inst{1} \and Robert H. Thomson\inst{2}  }

\authorrunning{R. Thomson}

\institute{Mount Holyoake College, South Hadley, USA \and Cognitive Security Institute\orcidID{0000-0001-9298-2870} \\ \email{robert.h.thomson@gmail.com}}

\maketitle

\begin{abstract}
Supervised machine learning techniques rely on labeled data to achieve high task performance, but this requires the labels to capture some meaningful differences in the underlying data structure. For training network intrusion detection algorithms, most datasets contain a series of attack classes and a single large benign class which captures all non-attack network traffic. A review of intrusion detection papers and guides that explicitly state their data preprocessing steps identified that the majority took the labeled categories of the dataset at face value when training their algorithms. The present paper evaluates the structure of benign traffic in several common intrusion detection datasets (NSL-KDD, UNSW-NB15, and CIC-IDS 2017) and determines whether there are meaningful sub-categories within this traffic which may improve overall multi-classification performance using common machine learning techniques. We present an overview of some unsupervised clustering techniques (e.g., HDBSCAN, Mean Shift Clustering) and show how they differentially cluster the benign traffic space. 

\end{abstract}


\section{Introduction}
Predominantly-used datasets for training intrusion detection algorithms include NSL-KDD \cite{tavallaee2009detailed}, UNSW-NB15\cite{moustafa2015unsw}, and CIC-IDS2017 \cite{sharafaldin2018toward}. Each of these pre-labeled datasets includes a set of attack classes and a singular \textit{normal} class representing all benign traffic. An informal review of numerous intrusion detection papers and guides identifying their preprocessing steps found that all but one \cite{alfoudi2022hyper} took the labeled categories of the dataset at face value when training their algorithms \cite{tavallaee2009detailed,moustafa2015unsw,moustafa2016evaluation,sharafaldin2018toward,wang2020explainable,dhanabal2015study,revathi2013detailed}. A question that has not been established is whether benign traffic is a relatively homogeneous class, or is it more heterogeneous? Given the nature of network flows coming from various services, ports, and sources, is it correct to lump all legitimate email, browsing, streaming, and downloading traffic into the same class? Are there potentially gaps in treating all benign traffic as a monolithic class which could allow attacks to be obfuscated between heterogeneous regions? The present paper answers these questions by conducting unsupervised clustering over benign traffic, reviewing the feature importance of these sub-classes to determine whether there are meaningful differences, and evaluating whether adding these sub-classes provides meaningful improvement in classification accuracy.

\subsection{Unsupervised Clustering Techniques}
\label{sec:clustering}
There are numerous techniques to do some form of unsupervised clustering and they generally fall into several functional types. The first type requires you to specify the number of categories in advance but the actual clustering is unsupervised. These techniques, such as K-Means and Gaussian Mixture Models \cite{kodinariya2013review} are best applied when you know in advance something about the nature of the data. K-Means specifically assumes that clusters are convex and similar-sized, which is not appropriate in intrusion detection datasets where there are substantial differences in category size both within attack categories and between attack categories and benign traffic. In addition, these methods are likely not appropriate as the underlying distribution of the benign traffic is precisely what we will be investigating. Conversely, there are unsupervised techniques which do not need to \textit{a priori} know the number of clusters, which includes Spectral Clustering, HDBSCAN (and the related OPTICS-DBSCAN and DBSCAN), and Mean Shift Clustering. While these techniques do not require a preset number of clusters, they are still parameterized to determine features such as the minimum cluster size, minimum density, as well as the choice of which distance function to use. Thus, it is still important to have some knowledge of the distribution of the data to ensure that the clusters created are meaningful to the task at hand (in this case, improving intrusion detection classification accuracy). Preliminary investigation revealed that spectral clustering is limited for larger datasets and assumes that the clusters should be similar-sized, so it was not pursued further \cite{von2007tutorial}. We thus considered HDBSCAN \cite{campello2013hdbscan} and Mean-Shift Clustering \cite{cheng1995mean} to cluster the intrusion detection datasets.

\subsubsection{Hierarchical Density-Based Spatial Clustering of Applications with Noise (HDBSCAN)}
HDBSCAN works by measuring the density of data points that surround each individual point, and then generating a hierarchy of clusters based on density thresholds (the \textit{min\_cluster\_size} parameter) \cite{campello2013hdbscan}. This technique is computationally complex ($\mathcal{O}(n \log n)$), but is able to find clusters of varying size and density. Specifically, the mutual reachability distance between two points \( x_i \) and \( x_j \) is defined as: 

\begin{equation}
{d_{mreach}(x_i, x_j) = \max \{ \text{core\_dist}(x_i),   \text{core\_dist}(x_j), d(x_i, x_j) \}}
\end{equation}

\noindent where \( d(x_i, x_j) \) is the actual distance between \( x_i \) and \( x_j \). 

The \textbf{core distance} of a point \( x \) is defined as: $\text{core\_dist}(x) = \max_{y \in N_k(x)} d(x, y)$ where \( N_k(x) \) is the set of the \( k \)-nearest neighbors of \( x \). Once the mutual reachability distance graph is completed, the minimum spanning tree of the graph is computed and single linkage clustering is performed to form a hierarchical tree. The trees are then condensed by removing branches with insufficient density.  

To select the final clusters, we define \textbf{cluster stability} as $S(C) = \sum_{x \in C} $ \\ $\int_{\lambda_{birth}}^{\lambda_{death}} d\lambda$ where \( \lambda \) is the inverse of the mutual reachability distance: \( \lambda = \frac{1}{d_{mreach}} \), and \( \lambda_{birth} \) and \( \lambda_{death} \) are the birth and death scales of the cluster. Clusters are selected by optimizing stability.

\subsubsection{Mean Shift Clustering}
Mean Shift clustering \cite{cheng1995mean} determines centroids using the density of nearby points, and then iteratively shifting nearby points toward those denser centers. Mean shift requires minimal assumptions about the underlying data and is effective in finding non-convex clusters, especially with large densities \cite{carreira2015meanshift}, although this comes at the expense of increased computational complexity ($\mathcal{O}(T n^2)$) where $T$ is the number of iterations and is ($\mathcal{O}(n \log n)$) in practice. The density function uses a kernel function: $f(x) = \sum_{i=1}^{n} K_h(x - x_i)$

\noindent where $K(x)$ is typically a Gaussian. To determine the higest-density regions, a gradient of the density estimate $f(x)$ is computed which represents the difference between the weighted mean of nearby points and the current position. Afterwards, each point $x$ is iteratively updated to move it towards a region of maximum density.

One drawback of Mean Shift Clustering is that it takes considerably longer to run compared to HDBSCAN, and can still be reliant on the bandwidth parameter which uses a distance metric to create a `window' of nearby points. As will be seen, the default clustering may produce many - potentially less-meaningful - clusters compared to HDBSCAN on relatively larger datasets.  

\subsection{Feature Importance}
The SHAP (SHapley Additive exPlanations) algorithm is a game theoretic measure to determine how much a given player contributes in a collaborative game. It has been adapted to machine learning applications to represent the average contribution of a given feature value to a model's prediction \cite{marcilio2020explanations,gaspar2024explainable}. The SHAP value for a specific feature in a particular instance is computed as the average marginal contribution of that feature across all possible \textit{coalitions} of features. This is done by considering all possible subsets of features and their corresponding predictions, and computing the difference in predictions when including the feature compared to when it is absent. This can also be used to show the relative feature importance \textbf{per class}, which is essential to determine whether there are meaningful differences between the potential sub-classes discovered by the techniques previously discussed in Section \ref{sec:clustering}. 

The SHAP value for feature $j$ and instance $i$ in a machine learning model can be computed using the following equation:
  \begin{equation}
      \phi_{j}^{i} = \sum_{S \subseteq \{1, 2, ..., M\} \setminus\{j\}} \frac{|S|!(M-|S|-1)!}{M!} [f(x_{i}^{S \cup \{j\}}) - f(x_{i}^{S})]    
  \end{equation}

\noindent where $\phi_{j}^{i}$ is the SHAP value for feature $j$ and instance $i$, $M$ is the total number of features, $S$ represents a subset of features excluding feature $j$, $x_{i}^{S}$ is the instance $i$ with features in subset $S$, and $f(x_{i}^{S \cup \{j\}})$ is the model's prediction for instance $i$ with feature $j$ added to subset $S$.

A specific variant for tree-based models like decision trees, random forest, and gradient boosted models is known as Tree SHAP, and is computed using $\phi_{j}^{i} = \sum_{p \in P_{i}^{j}} \frac{1}{|P_{i}^{j}|} \cdot (v_{p}^{j} - v_{\emptyset}^{j})$ where $\phi_{j}^{i}$ is the SHAP value for feature $j$ and instance $i$, $P_{i}^{j}$ is the set of paths in the tree that lead to instance $i$ where feature $j$ is active, $|P_{i}^{j}|$ is the number of paths in $P_{i}^{j}$, $v_{p}^{j}$ is the prediction value of feature $j$ for path $p$, and $v_{\emptyset}^{j}$ is the expected prediction value of feature $j$ across all instances. This equation captures the contribution of feature $j$ to the prediction for instance $i$ by considering all possible paths that reach $i$ where $j$ is active, and comparing the feature value along each path to the expected feature value.

\subsection{Datasets}
For the present study we examined three traditionally-used datasets described fully below: NSL-KDD, UNSW-NB15, and CIC-IDS 2017. Each of these network intrusion datasets consists of a set of instances with a mix of normal traffic with several kinds of malicious attacks interspersed. Because the purpose of these datasets is training, there is a relatively higher proportion of attacks than would be seen in real network traffic flows. 

\subsubsection{NSL-KDD}
NSL-KDD \cite{tavallaee2009detailed} is an updated and cleaned version of the popular KDD-Cup '99 \cite{revathi2013detailed} that addresses several concerns with the original dataset, namely removing redundant records and creating a scalable dataset where common machine learning techniques could be implemented and executed on a single machine. The dataset contains predefined training and testing sets with 148,517 total instances (77,054 benign), and 41 separate features. Of those features, 21 refer to the external connection and 19 describe connections within the host with one meta-feature. \cite{dhanabal2015study} examined feature importance for NSL-KDD models and \cite{wang2020explainable} found substantial agreement between SHAP values LIME. 
 
\subsubsection{UNSW-NB15}
UNSW-NB15 contains 9 forms of malicious traffic across 2,504,044 instances and 49 features \cite{moustafa2015unsw}. Of this data, 2,218,761 rows are considered Normal and 321,283 are malicious. The team behind UNSW-NB15 also developed a standardized evaluation set with 175,341 (93,005 benign) training and 82,232 (37,000 benign) testing instances with 44 features. 

\subsubsection{CIC-IDS 2017}
CIC-IDS 2017 is another network traffic data set, created by the Canadian Institute for Cybersecurity \cite{panigrahi2018detailed}. The full data set contains 2,830,742 instances of network traffic, which contains 2,273,097 instances of traffic labeled ”BENIGN” and 557,646 instances of malicious traffic divided into 14 categories with 80 features.  


\subsection{Research Questions}
As previously discussed, the majority of extant research using common intrusion detection datasets assumes that the labels are correct and meaningful. In the case of specific attack categories, their labels are grounded as being representative of a specific known attack. Conversely, in the case of benign traffic it is essentially all traffic which isn't an attack. There is an outstanding question whether this data is relatively homogeneous or heterogeneous, and if heterogeneous, whether identifying meaningful sub-classes will improve classification performance. Based on this two-part underlying question, we make the following hypotheses:

\begin{enumerate}
    \item Benign traffic in each of the datasets will be heterogeneous.
    \item The sub-classes will contain meaningful content reflecting increased performance by models using clustered benign traffic. 
\end{enumerate}

\section{Methods}

\subsection{Preprocessing Steps}
\label{sec:preprocessing}
Some common steps were used across all three datasets. A Python Pipeline was created which ran all categorical variables through \textit{OneHotEncoder} and numeric variables were all scaled using  scikit-learn's \textit{MinMaxScaler()}. The NSL-KDD and UNSW-NB15 datasets were already cleaned of missing and incomplete values, with only one feature in the UNSW-NB15 dataset requiring correction. In the UNSW-NB15 data, all values greater than 1 were set to 1 in the \textit{is-ftp-login} binary feature. In the final step, \textit{ColumnTransformer} was applied to ensure the same number of features was in the training and testing subsets of each dataset.

CIC-IDS 2017 was relatively unprocessed and required several preprocessing steps. Duplicate rows and those with NaN or infinite values were dropped from the data set. Moreover, rows containing negative values in the features of \textit{Flow Duration}, \textit{Flow Bytes/s}, \textit{Flow IAT Mean}, \textit{Fwd Header Length}, and \textit{Bwd Header Length} were removed since negative values are not consistent given the nature of these features. Features containing solely 0 values were deemed uninformative and subsequently removed, encompassing items like \textit{Fwd Avg Bytes/Bulk}, \textit{Bwd PSH Flags,} and \textit{Bwd URG Flags}. Features irrelevant to the experimentation scope, such as \textit{Destination Port} were also discarded. \textit{Idle Mean}, \textit{Idle Std}, \textit{Idle Max}, and \textit{Idle Min} were also removed due to their excessive standard deviations. Lastly, with respect to feature selection, the removal of \textit{Init win bytes fwd} was informed by \cite {pelletier2020evaluating} who identified it as a strong predictor for the \textit{Label} column. 

This led to a refined data set with the removal of approximately 310,000 instances (of 2.8 million) and a reduction down to 64 features (from 80). Of that, the subset of 173k was chosen to match the size of the UNSW test set and to keep things computable on a single machine. Sampling to 56k benign and 173k total rows was accomplished using the Pandas \textit{.sample()} function. The attack distribution of the original uncleaned data set marginally diverges from the cleaned distribution due to certain attack classes (namely \textit{Portscan} and \textit{SSH Patator}) exhibiting relatively more instances being removed during cleaning when compared to other attack types.

\subsection{Machine Learning Classification}
HDBSCAN and Mean Shift Clustering was performed on each of the three datasets' benign traffic. Numerous pilot runs were performed on HDBSCAN to understand explore the \textit{minimum sample size} and \textit{minimum cluster size} parameters (values included 2,5,10,15,25,50,10,250,500,1000 \& 10000 for each attribute). Smaller sizes led to exponentially more sub-clusters for benign traffic. The values reported in this paper are for HDBSCAN with min\_sample of 500 and min\_cluster\_size of 10000, which generally provided 3-6 clusters for the UNSW-NB15 and NSL-KDD dataset\cite{khraisat2019survey}. Mean Shift Clustering was run with default parameters. 

The datasets were re-integrated and RandomForest models were fit to each of the base dataset, the HDBSCAN clustered data, and the Mean Shift clustered data. In total 9 models were fit. Models were evaluated based on accuracy, precision, recall, and F-1 score. Ground truth was assessed for 3 different candidate solutions: 1) the \textit{base} model which trained the classifier on the default benign traffic, 2) the \textit{benign-clustered} model which trained the classifier using the clusters of the benign traffic generated by HDBSCAN and Mean Shift Clustering, and 3) the \textit{benign-rejoined} model which evaluated the classifier by taking the clustered ground truth (e.g., normal-1, normal-2...) and converting back to a single benign category. This third case would evaluate a more comparable model to the base model given we did not intend for a normal cluster being misclassified as another cluster to be necessarily considered an error.

\section{Results}
These results describe the differences in sub-clusters from benign traffic in the UNSW-NB15, NSL-KDD, and CIC-IDS 2017 datasets. In this section, we will describe the number of clusters presented, model performance, consistency between methods, and utilize SHAP values to determine whether the clusters present any meaningful difference in network traffic. Table \ref{tab:ARI} presents the number of sub-clusters and the agreement between clusters and methods as measured my adjusted mutual information and adjusted Rand score. Overall, despite Mean Shift generating many more clusters, there is relatively high agreement between methods for all three datasets. 

\begin{table}[htbp]
    \centering
    \begin{tabular}{|c|c|c|c|}
    \hline
         & \textbf{UNSW} & \textbf{NSL} & \textbf{CICIDS}  \\
    \hline

         ARS & .37 & .68 & .60 \\
         AMI & .60 & .57 & .66 \\
         \hline
         HDBSCAN & 4 & 3 & 6 \\
         MS & 5 & 75 & 35 \\
         \hline
    \end{tabular}
    \caption{Adjusted Rand Score (ARS) and Adjusted Mutual Information Score (AMI) comparison between HDBSCAN and Mean Shift Clustering, including the number of sub-clusters. Numbers approaching 1 reflect perfect agreement and 0 reflects no agreement. 
    HDBSCAN has some amount of instances end up in an 'unclustered' class; while Mean Shift clustering puts outlier values into their own cluster, reflecting a long tail of clusters with <.01\% of the data. Overall clusters are relatively similar between methods.}
    \label{tab:ARI}
\end{table}

\begin{figure}[htbp]
    \centering
    
    \begin{subfigure}[b]{1\textwidth}
        \centering
        \includegraphics[width=0.3\textwidth]{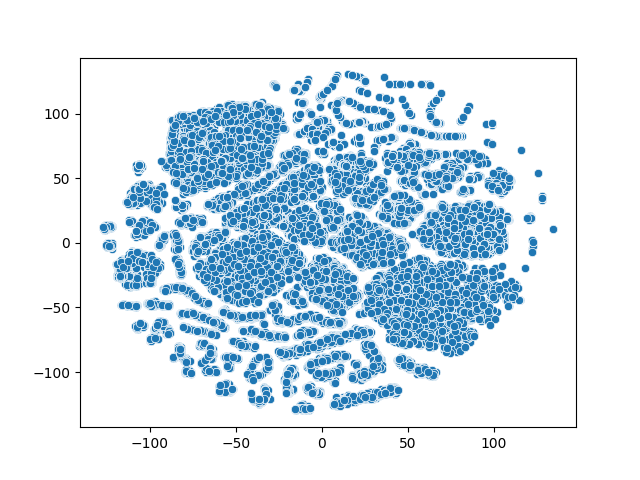}
        \includegraphics[width=0.3\textwidth]{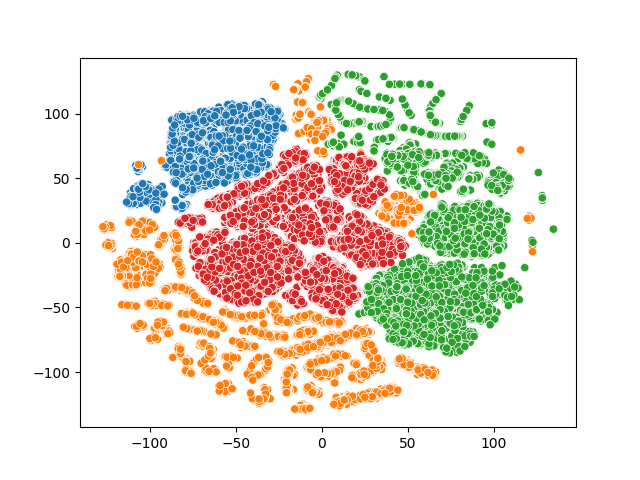}
        \includegraphics[width=0.3\textwidth]{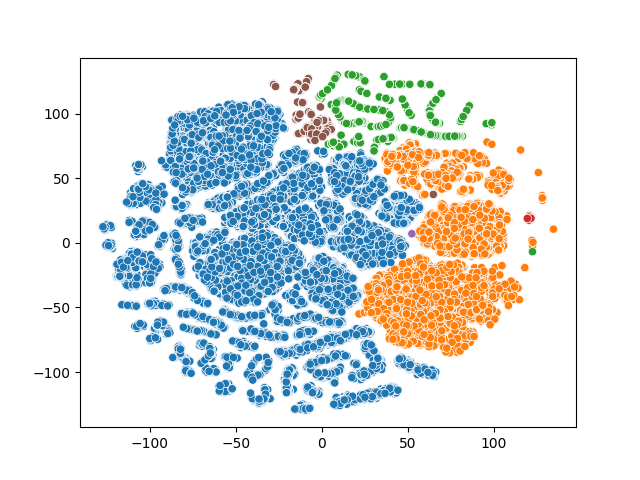}
        \caption{UNSW-NB15}
        \label{fig-tsne-unsw}
    \end{subfigure}
    
    \begin{subfigure}[b]{1\textwidth}
        \centering
        \includegraphics[width=0.3\textwidth]{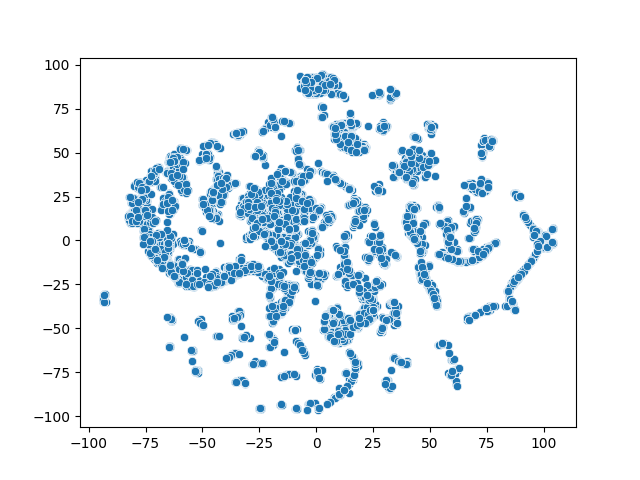}
        \includegraphics[width=0.3\textwidth]{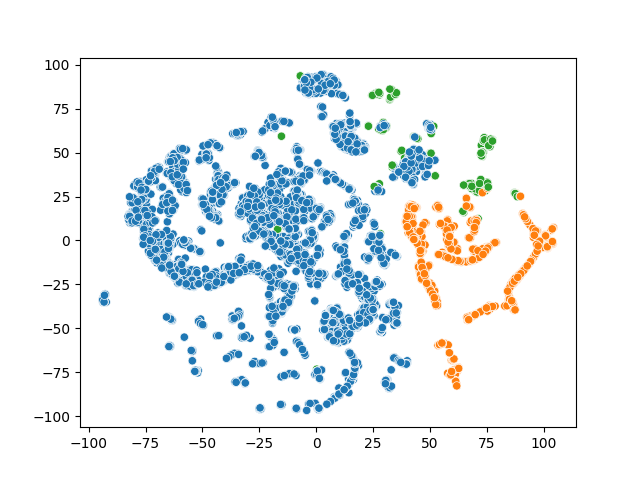}
        \includegraphics[width=0.3\textwidth]{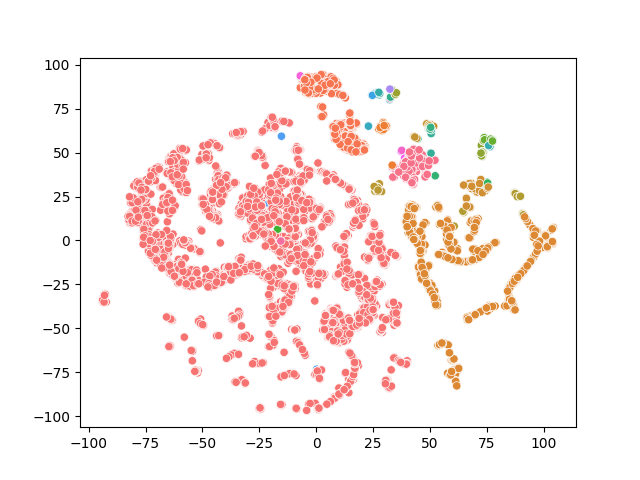}
        \caption{NSL-KDD}
        \label{fig-tsne-nsl}
    \end{subfigure}

    \begin{subfigure}[b]{1\textwidth}
        \centering
        \includegraphics[width=0.3\textwidth]{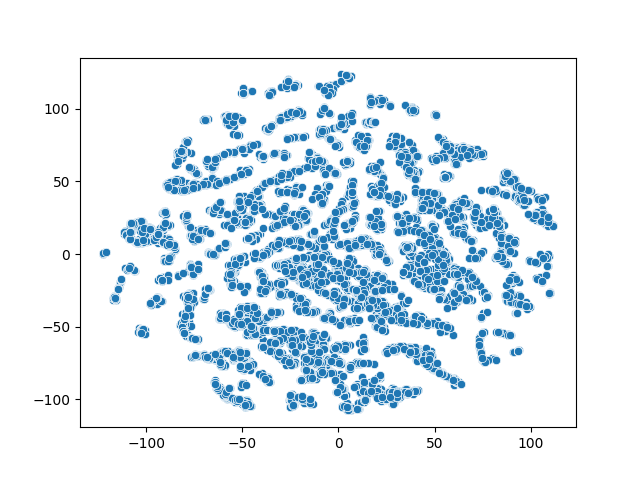}
        \includegraphics[width=0.3\textwidth]{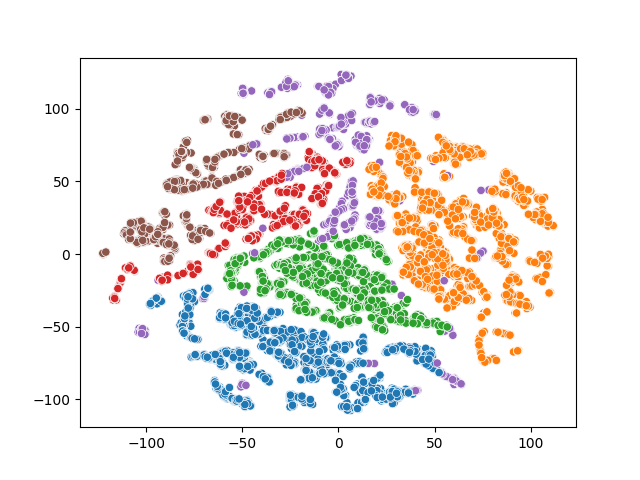}
        \includegraphics[width=0.3\textwidth]{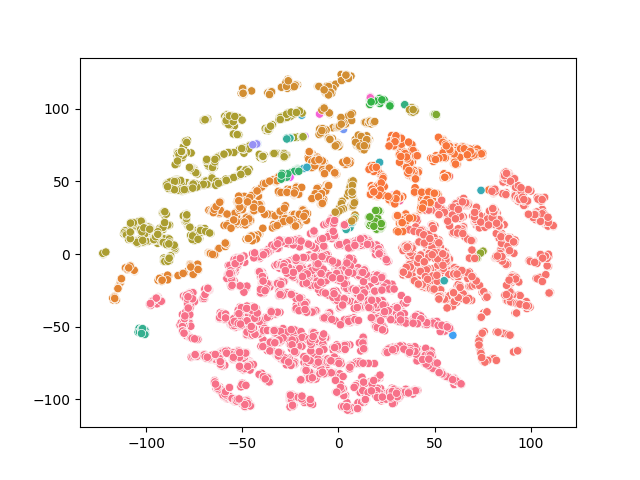}
        \caption{CIC-IDS 2017}
        \label{fig-tsne-cicids}
    \end{subfigure}

    \caption{The panels present t-SNE visualizations for each dataset. The plot on the left is the base t-SNE visualization, with HDBSCAN output in the center plot and Mean Shift clustering on the right plot. The colors reflect different clusters and are not consistent between HDBSCAN and Mean Shift models. It is not possible to directly align these clusters although they do have an overall high agreement as per Table \ref{tab:ARI}.}
    \label{fig:tsne}
\end{figure}

To visualize these clusters, we employed t-distributed stochastic neighbor embedding (t-SNE) and superimposed the HDBSCAN and Mean Shift clusters onto the plot (see Figure \ref{fig:tsne}). t-SNE is already an unsupervised dimensionality-reduction technique which may display some patterns in data, although the interpretation of those patterns is entirely up to the user. By superimposing HBBSCAN and Mean Shift sub-clusters onto the data, it provides some measure of consistency.  Visually it does appear that the sub-clustered generated by HDBSCAN and Mean Shift clustering do map onto consistent regions consolidated by t-SNE, which implies that there is some consistency in the underlying benign data which all clustering algorithms are consistently finding. 

\begin{figure}[htbp]
    \centering
    
    \begin{subfigure}[b]{1\textwidth}
        \centering
        \includegraphics[height=105px]{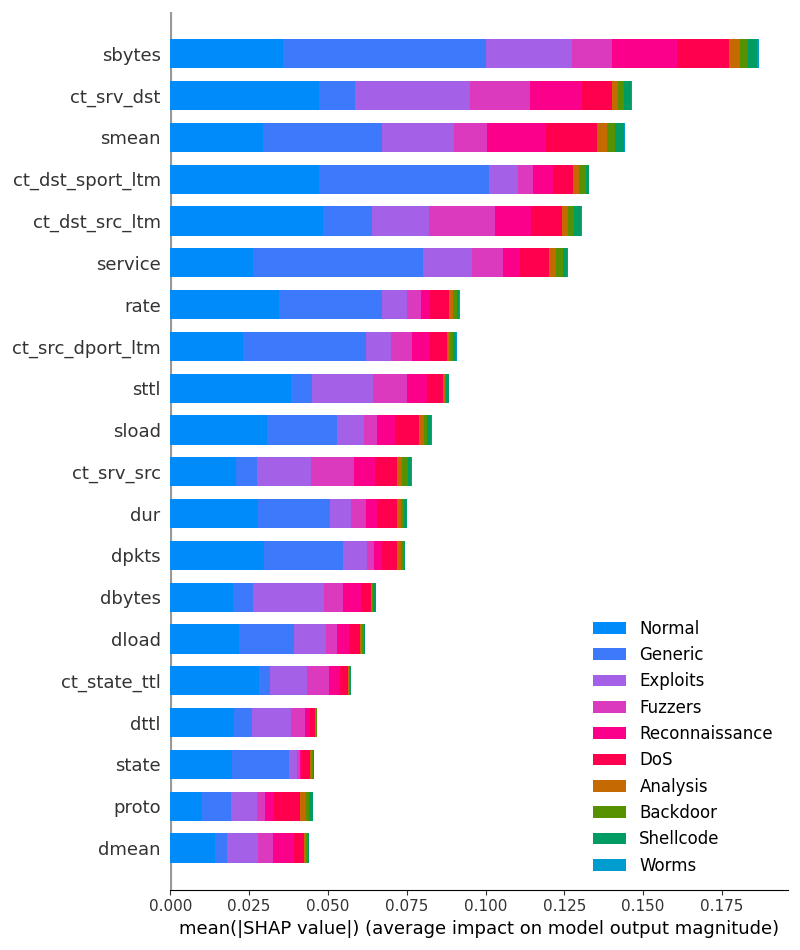}
        \includegraphics[width=0.3\textwidth]{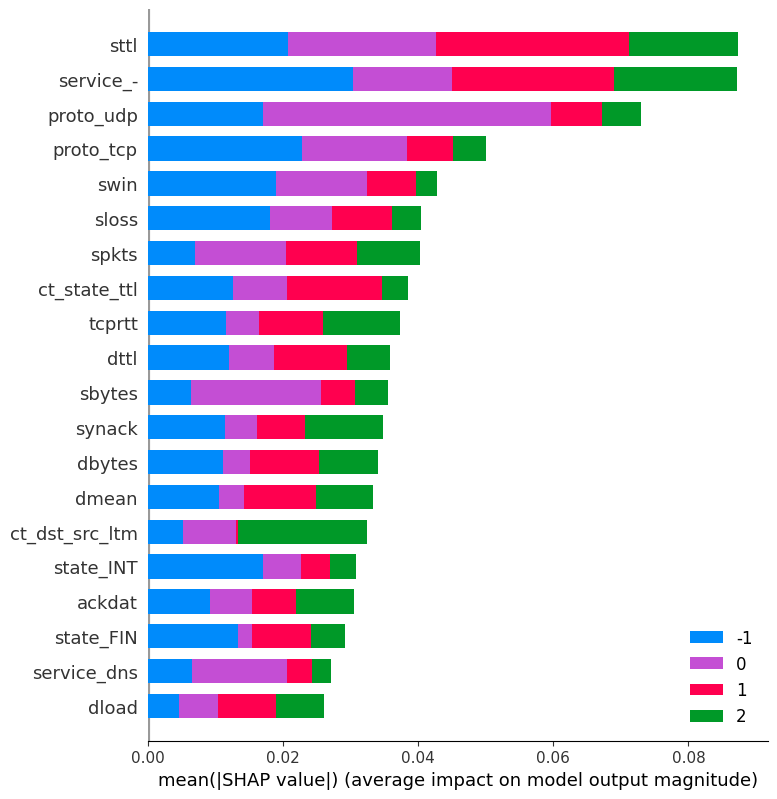}
        \includegraphics[width=0.3\textwidth]{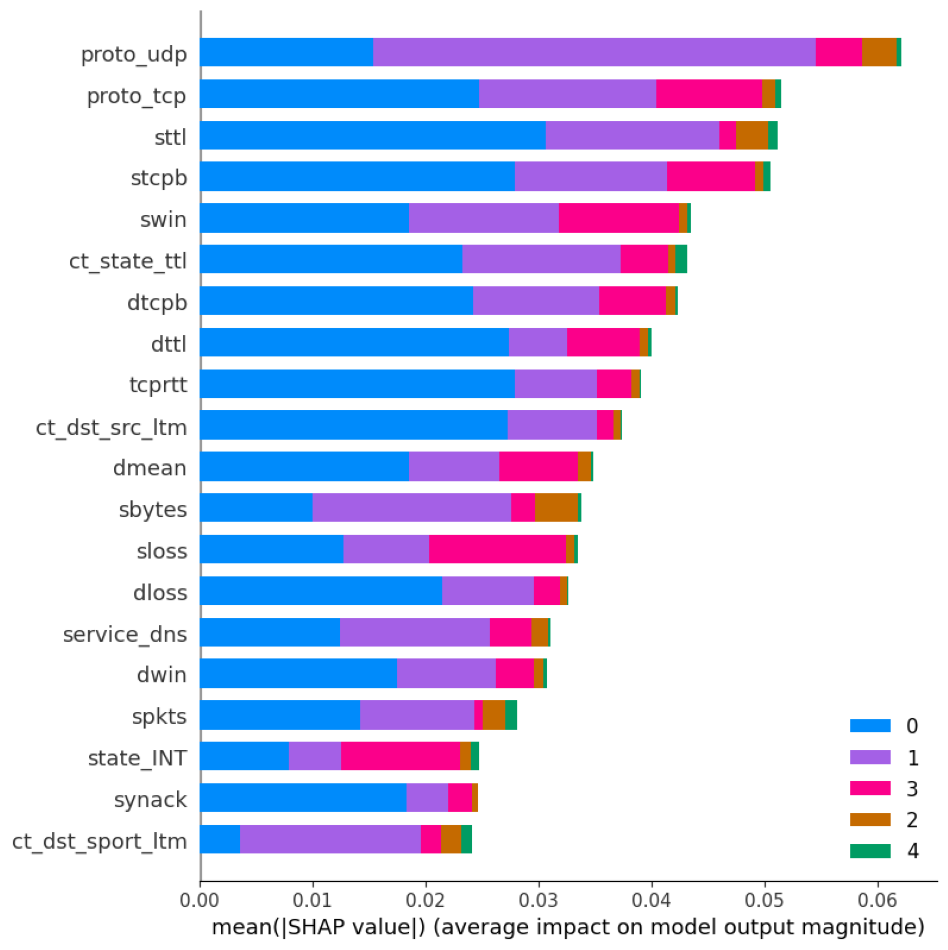}
        \caption{UNSW-NB15}
        \label{fig-shap-unsw}
    \end{subfigure}
    
    \begin{subfigure}[b]{1\textwidth}
        \centering
        \includegraphics[height=95px]{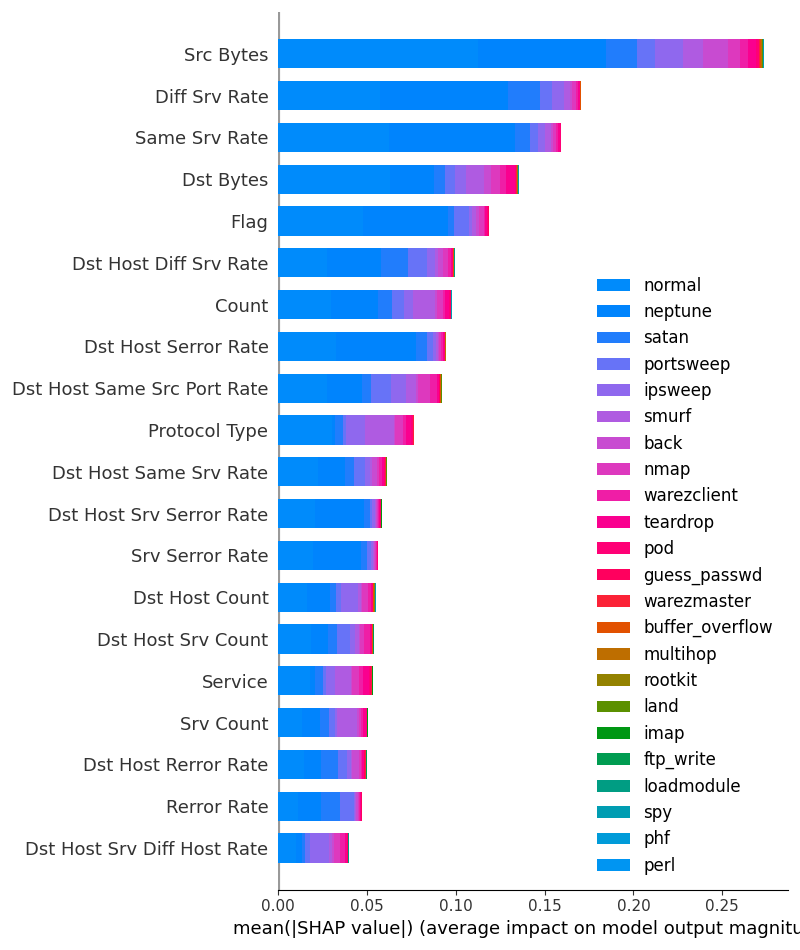}
        \includegraphics[width=0.3\textwidth]{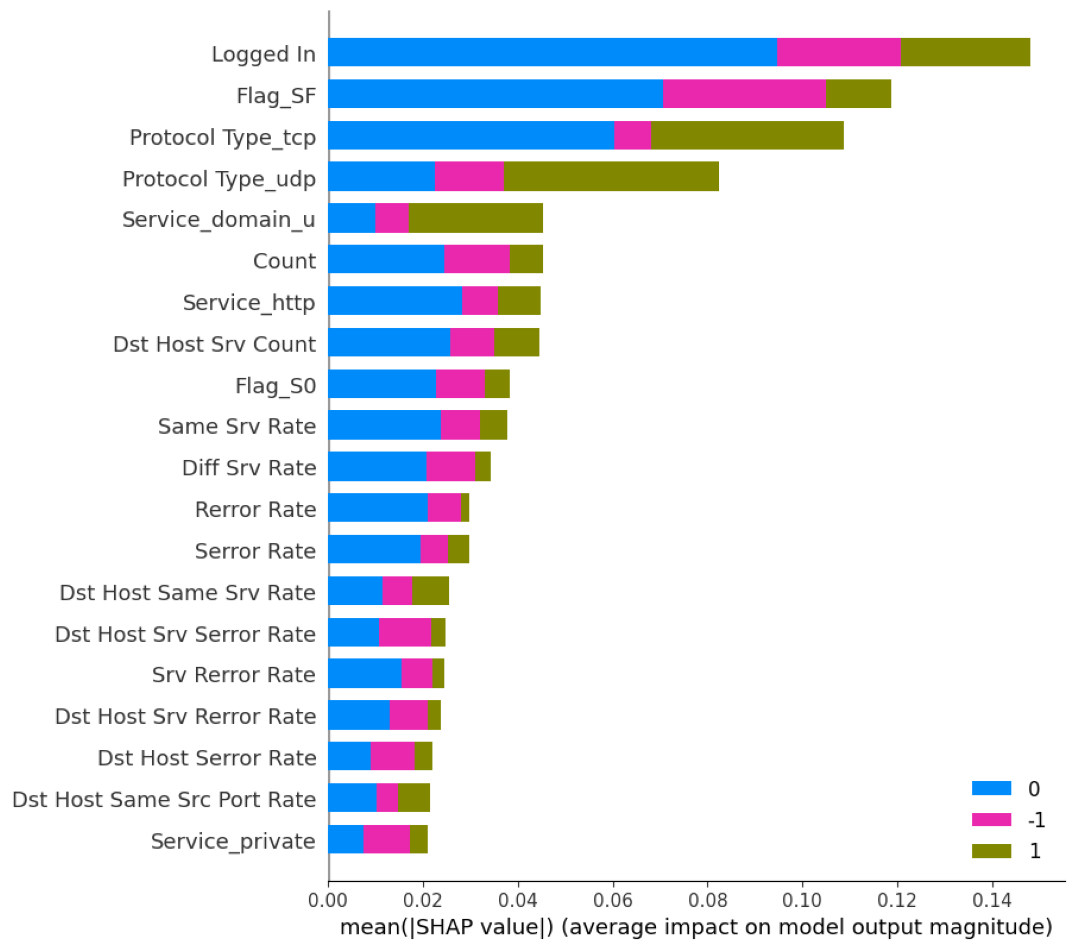}
        \includegraphics[width=0.3\textwidth]{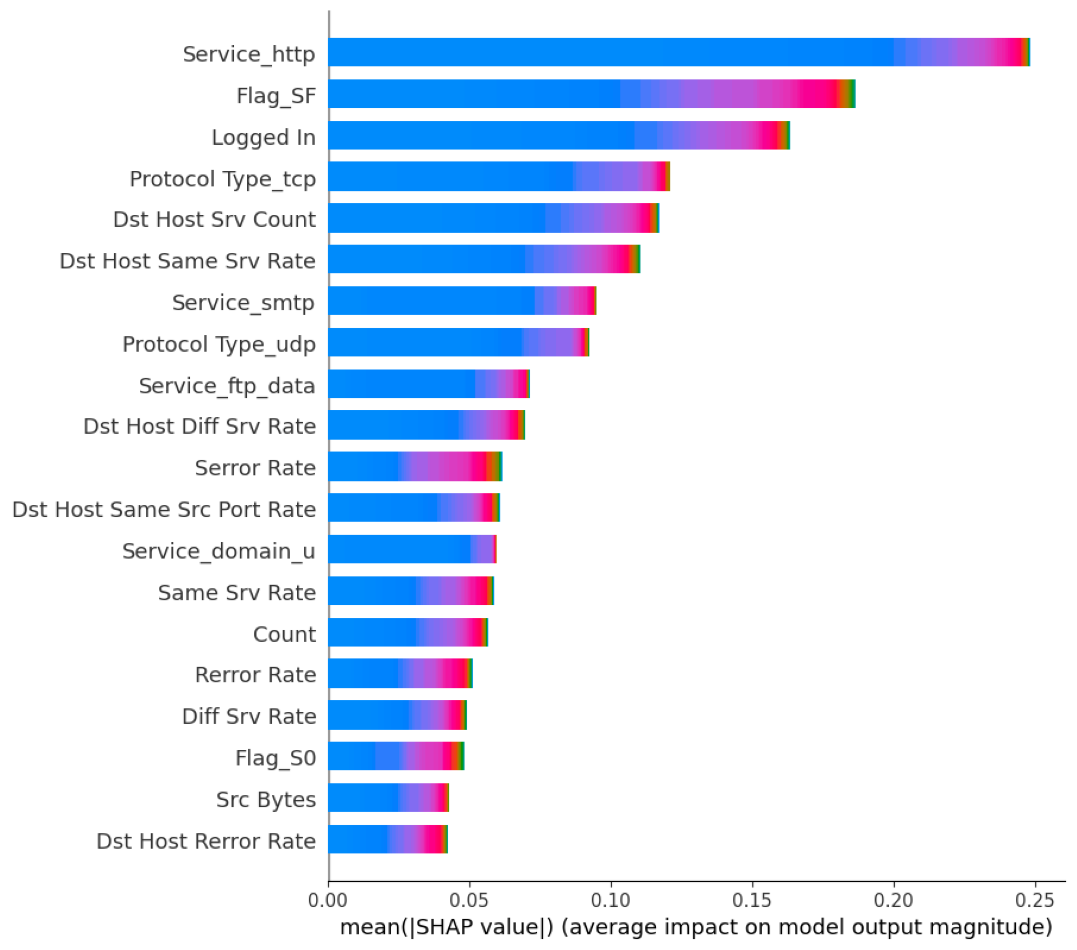}
        \caption{NSL-KDD}
        \label{fig-shap-nsl}
    \end{subfigure}

    \begin{subfigure}[b]{1\textwidth}
        \centering
        \includegraphics[height=95px]{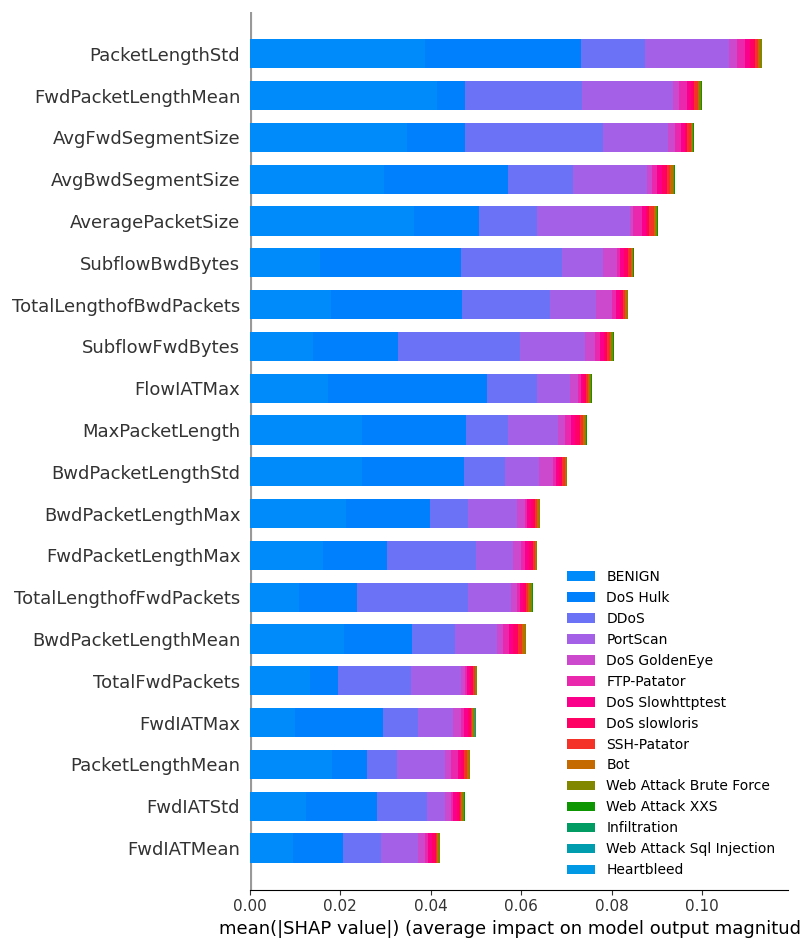}
        \includegraphics[width=0.3\textwidth]{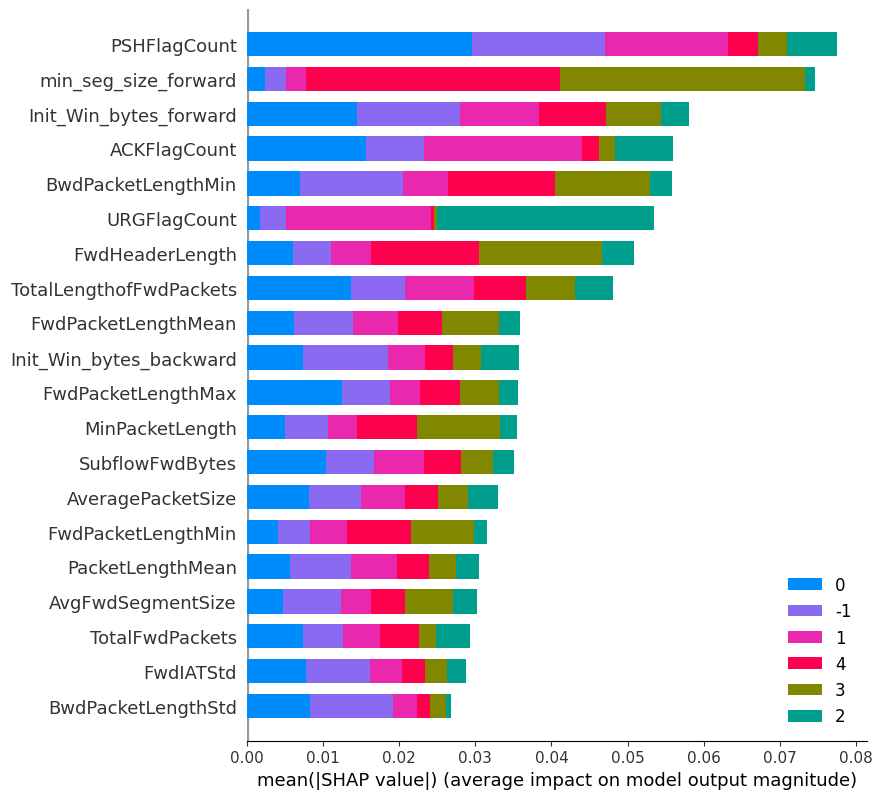}
        \includegraphics[width=0.3\textwidth]{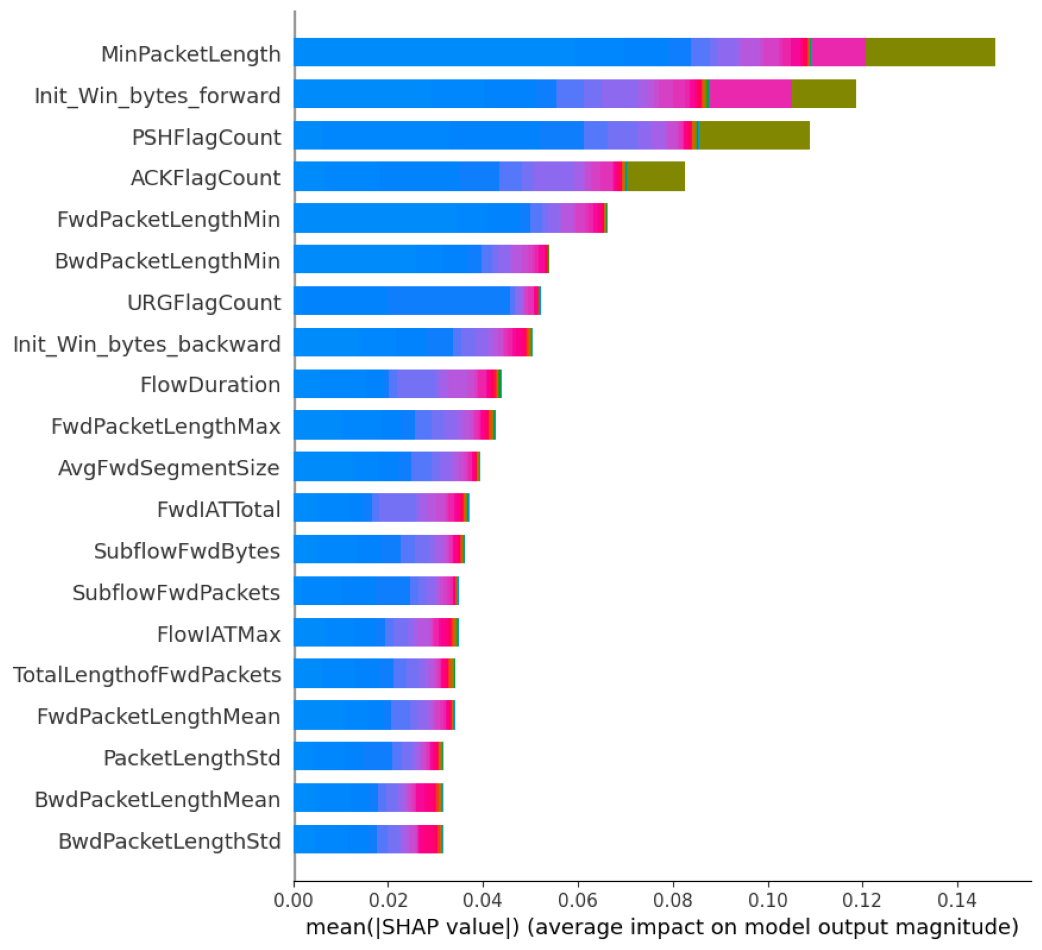}
        \caption{CIC-IDS 2017}
        \label{fig-shap-cicids}
    \end{subfigure}

    \begin{subfigure}[b]{1\textwidth}
        \centering
     \includegraphics[width=0.36\textwidth]{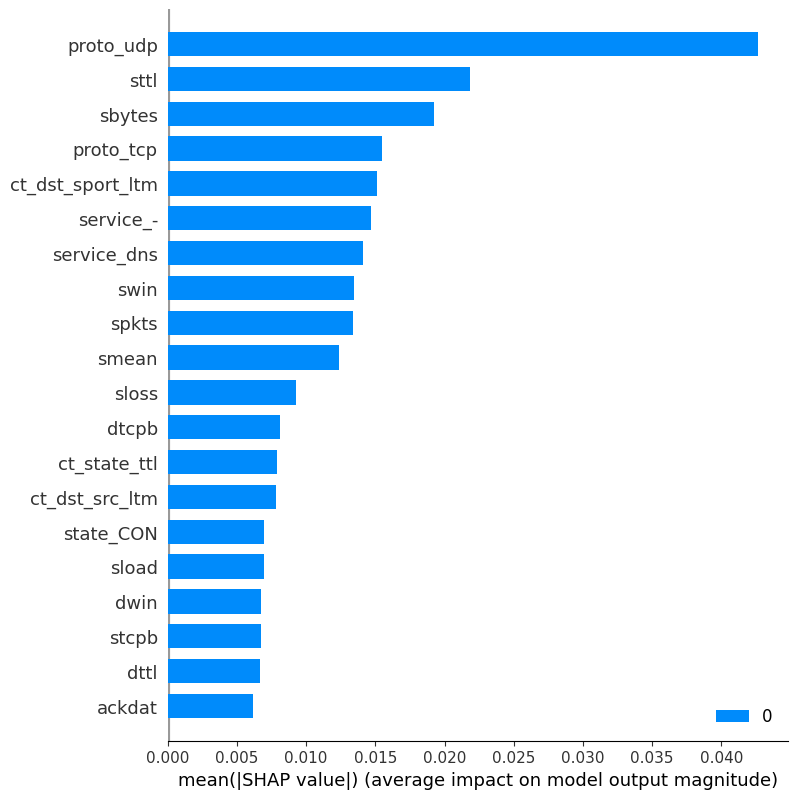}
     \includegraphics[width=0.33\textwidth]{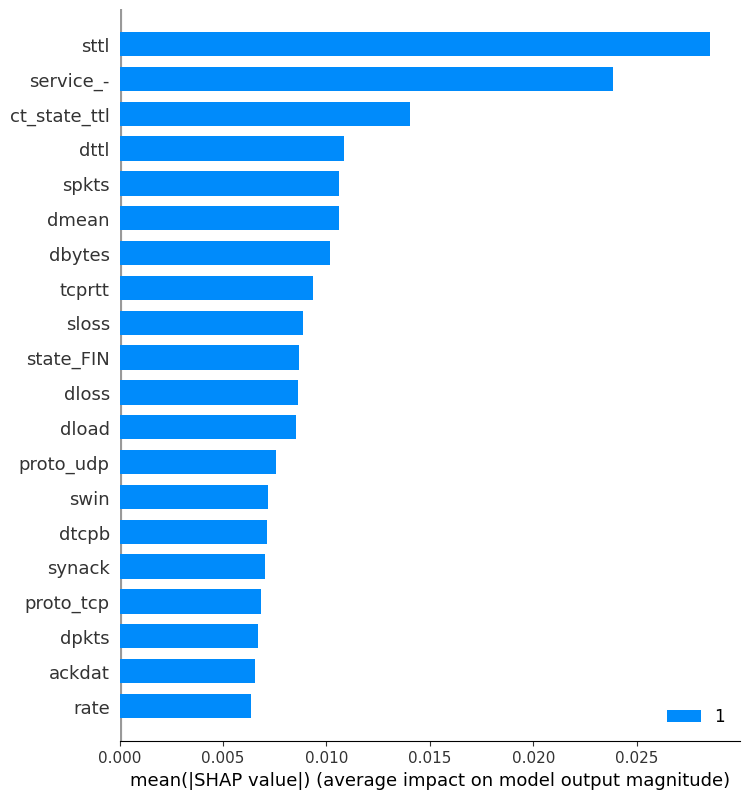}
     \caption{SHAPley values between HDBSCAN cluster 0 and 1 in UNSW-NB15.}
     \label{fig:sub-cluster}
    \end{subfigure}

    \caption{SHAPley values for each dataset. The left plot includes benign traffic and attack categories. The plot in the center panel is the relative importance of each sub-cluster for HDBSCAN, while the right plot is the relative importance for Mean Shift Clustering. As is most apparent in UNSW-NB15, there are different relative importance of each feature in the sub-clusters.}
    \label{fig:SHAP}
\end{figure}

Investigating SHAP values for each category (see Figure \ref{fig:SHAP} further highlights that the different clusters have some apparently meaningful differences. For instance, in UNSW-NB15 with HDBSCAN, the model presents different importance to the STTL and proto\_udp features compared between sub-cluster 0 and sub-cluster 1 (see Figure \ref{fig:sub-cluster}). Rank Biased Ordering (RBO) \cite{webber2010similarity} is a technique to determine whether two lists are in agreement, and is often used when comparing the relative result of two orderings such as comparative output on search engines or other recommender systems. The  $RBO_{MIN}$ was .34 and the $RBO_{EXT}$ was .42 with a $p = .95$ reflecting the top-10 features having 67\% of the weight in the ranking. RBO values range from 0 (maximally dissimilar) to 1 (maximally similar) reflecting that these categories are substantially dissimilar. These results imply that the sub-clusters do contain some meaningful differences, supporting our first hypothesis that benign traffic has heterogeneous regions, at least in the three intrusion detection datasets. The second question, however, is whether explicitly labeling this heterogeneity improves performance?

Unfortunately, our second hypothesis was not supported as the models do not significantly differ in their performance using clustered benign traffic.  Accuracy and F1-Score was unchanged  (86.8\% Accuracy and 86.2\% F1-Score for both the Base model and Clustered Normal for UNSW; 72.1\% Accuracy for both with 82.3\% vs 80.4\% F1-Score for Base vs Clustered Normal for NSL-KDD, and 99.6\% Accuracy and F1-Score for Base and Clustered Normal for CIC-IDS 2017. While outside the scope of the present paper, confusion-matrices also show no difference in individual category performance between models. 


\section{Discussion}
Our hypotheses received partial support in that there does appear to be some meaningful heterogeneity in each of the datasets as measured by multiple unsupervised categorization techniques, however, explicitly labeling this heterogeneity does not improve performance. A challenge is that a user requires some domain expertise to determine sample and cluster sizes without the requirement to conduct substantial parameter exploration.

A caveat with the results presented in the CIC-IDS 2017 dataset is that the performance is near-ceiling level. Noteworthy studies by others \cite{pelletier2020evaluating,sharafaldin2018toward} have also reported remarkably high model performance scores. This high performance possibly suggests the presence of extraneous variables that might facilitate the classifiers in accurately identifying various attack categories. When developing the CIC-IDS 2017 data set, the authors initiated specific attack types on specific days and times. It is plausible that several of the features associated with different attack categories are closely correlated with time.

\subsubsection{Acknowledgments} 
This research was sponsored by the ONR MURI Grant Number W911NF-17-1-0370 as well as C5ISR agreement USMA23011. The views contained therein are those of the authors and should not be interpreted as representing the official policies, either expressed or implied, of the U.S. Government.

\bibliographystyle{splncs04}
\bibliography{references} 

\end{document}